\documentclass[epj]{webofc}

\usepackage[utf8]{inputenc}
\usepackage{soul}
\usepackage[varg]{txfonts}   
\usepackage{booktabs}
\usepackage{xcolor}
\definecolor{darkred}{rgb}{0.4,0.0,0.0}
\definecolor{darkgreen}{rgb}{0.0,0.4,0.0}
\definecolor{darkblue}{rgb}{0.0,0.0,0.4}
\usepackage[bookmarks,linktocpage,colorlinks,
    linkcolor = darkred,
    urlcolor  = darkblue,
    citecolor = darkgreen]{hyperref}
    
\usepackage{amsfonts}
\usepackage{amssymb}
\usepackage{tikz}

\usepackage{soul} 

\usepackage{subfigure}
\wocname{EPJ Web of Conferences}
\woctitle{Lattice2017}
\begin{document}
%
\selectlanguage{english}
\title{%
Improved convergence of Complex Langevin simulations
}
\author{%
\firstname{Felipe} \lastname{Attanasio}\inst{1,2} \and
\firstname{Benjamin}  \lastname{Jäger}\inst{3,4}\fnsep\thanks{Speaker,
\email{jaeger@imada.sdu.dk}}
}
\institute{%
Department of Physics, University of Washington, Box 351560, Seattle, WA 98195, USA
\and
Department of Physics, College of Science, Swansea University, Swansea SA2 8PP, United Kingdom
\and
CP3-Origins \& Danish Institute for Advanced Study, Department of Mathematics and Computer Science,
University of Southern Denmark, 5230 Odense M, Denmark
\and
Institute for Theoretical Physics, ETH Zürich, CH-8093 Zürich, Switzerland
}
\abstract{%
The sign problem appears in lattice QCD as soon as a non-zero chemical potential is
introduced. This prevents direct simulations to determine the phase structure of the strongly
interacting matter. Complex Langevin methods have been successfully used for various models
or approximations of QCD. However, in some scenarios it converges to incorrect
results. We present developments of our new method that helps to improve the convergence 
by keeping the system closer to the SU$(3)$ manifold and discuss preliminary
tests and results.
}
\maketitle
\section{Introduction}\label{intro}

{\em Sign problems} have been found, and obstructed
progress, in many areas of physics. One of the most famous examples is QCD with
non-vanishing Baryon chemical potential in Euclidean spacetime.
Here the fermion determinant, which appears after integrating
out the fermionic degrees of freedom, becomes complex and thus prevents 
direct simulation via Monte Carlo methods.
Complex Langevin simulations offer an alternative to study complex-valued Euclidean path integrals.
It has been shown that Complex Langevin methods are applicable even when the sign problem is severe~\cite{Aarts:2008rr, Sexty:2013ica, Aarts:2014bwa,Sinclair:2015kva,Nagata:2016vkn}.
The technique is based on the principle of stochastic quantisation and has been proposed decades ago by Klauder and Parisi~\cite{Parisi:1980ys, Parisi:1984cs, Klauder:1983nn, Klauder:1983zm, Klauder:1983sp}.
More recently, the development of {\em gauge
cooling}~\cite{Seiler:2012wz,Aarts:2013uxa} has enabled complex Langevin simulations of QCD with heavy
quarks~\cite{Aarts:2015yba,Aarts:2016qrv} and light quarks~\cite{Sexty:2013ica,
Aarts:2014bwa,Aarts:2014fsa}. Nevertheless, simulations with smaller gauge
couplings, typically below $\beta\sim 5.8$, do not converge to the
correct results. To tackle this issue we have introduced an additional force, 
named Dynamic Stabilisation~\cite{Aarts:2016qhx}, 
which is expected to vanish in the continuum limit.
In the following we present an update on our studies of dynamic stabilisation 
for QCD in the limit of heavy quarks, and apply it also to the XY model at
finite chemical potential~\cite{Aarts:2010aq}.

\section{Complex Langevin simulation and Dynamic Stabilisation}\label{CLE}

The (remaining) degrees of freedoms after integrating the fermions out, i.e. the SU($3$) gauge links $U_{x,\mu}$, are
evolved for a small time step $\varepsilon$ along a fictitious time dimension, known as the Langevin time $\theta$, using~\cite{Damgaard:1987rr}
\begin{equation}
	U_{x,\mu}(\theta + \varepsilon) = \exp \left[ \mathrm{i} \, \lambda^a \left( 
	\varepsilon \, K_x^a  + \sqrt{\varepsilon} \,
	\eta^a_{x,\mu} \right) \right]\, U_{x,\mu}(\theta),
	\label{eq:CLE}
\end{equation} 
where $\lambda^a$ are the Gell-Mann matrices, $\eta^a_{x,\mu}$ are white 
noise fields and the drift $K^a_{x,\mu}$ is given by
\begin{equation}
K_{x,\mu}^a = - D^a_{x,\mu} \, S,  \quad  \text{where} \quad  D^a_{x,\mu} f(U) = \left.\frac{\partial}{\partial \alpha} f(e^{i \alpha \lambda^a} U) \right|_{\alpha=0}\,.
\end{equation}
If the drift becomes complex, the Langevin equation naturally extends into the
larger gauge group of SL$(3,\mathbb{C})$. The action $S$ contains the Yang-Mills
plaquette action for the gluons and the fermion determinant $\det D$,
representing the quarks degrees of freedom. The effective action can be written as 
\begin{equation}
	S = S_{YM} - \ln \det D.
\end{equation} 
For simplicity, we ignore issues arising from meromorphic drifts
originating from the logarithm of the fermion determinant, but refer the reader
to the discussions
in~\cite{Mollgaard:2013qra,Splittorff:2014zca,Nishimura:2015pba,Greensite:2014cxa,Aarts:2017vrv}.
The fact that SL($3,\mathbb{C}$) is not a compact group allows for runaway trajectories.
This is usually monitored by the so-called unitarity norm,
\begin{equation}
d = \frac{1}{3\,V} \sum_{x, \mu} \mathrm{Tr}\left( U_{x,\mu} U_{x,\mu}^\dagger
-1 \right)^2\,,
\end{equation}
where $V$ is the lattice $4-$volume, which measures the distance to the SU($3$) manifold.
It is known that large unitarity norms lead to incorrect results~\cite{Aarts:2016qrv}.
Gauge cooling has been constructed to keep $d$ small, but is ineffective in some situations.
As proposed in~\cite{Aarts:2016qhx,Attanasio:2016mhc}, we modify the Langevin
drift
\begin{equation}
K_{x,\mu}^a \to - D^a_{x,\mu} \, S + \textcolor{darkred}{\mathrm{i} \,\alpha_{\mathrm{DS}}\, M^{a}_{x}}
\end{equation}
by adding a term that only acts in the non-SU$(3)$ directions and has a restoring character, i.e.
it will reduce the distance to the SU$(3)$ manifold. A possible choice for the force is given by
\begin{equation}
M^a_x = i \, b^a_x \, \Big( \sum_c b^c_x\, b^c_x \Big)^3, \quad \text{where}
\quad b^a_x =
	\mathrm{Tr}
	\Big[ \lambda^a \sum_\nu U_{x,\nu} U^\dagger_{x,\nu} \Big].
\end{equation}
The ``strength'' of this force can be changed by modifying the control parameter
$\alpha_{\mathrm{DS}}$. By construction, the force $M_x^a$ grows rapidly with the
distance to the SU$(3)$ manifold, $M \sim d^7$.
Since it has been previously shown that $d$ should be kept significantly smaller
than one ($d\sim0.03$ has been taken as a conservative threshold) ~\cite{Aarts:2016qrv}, only sufficiently large control
parameters have a non-trivial effect. It is important to note that DS will, for
very large $\alpha_{\mathrm{DS}}$, effectively re-unitarise the theory and thus produce
incorrect results.
\begin{figure}[thb]
  \centering
  \includegraphics[width=9cm,clip]{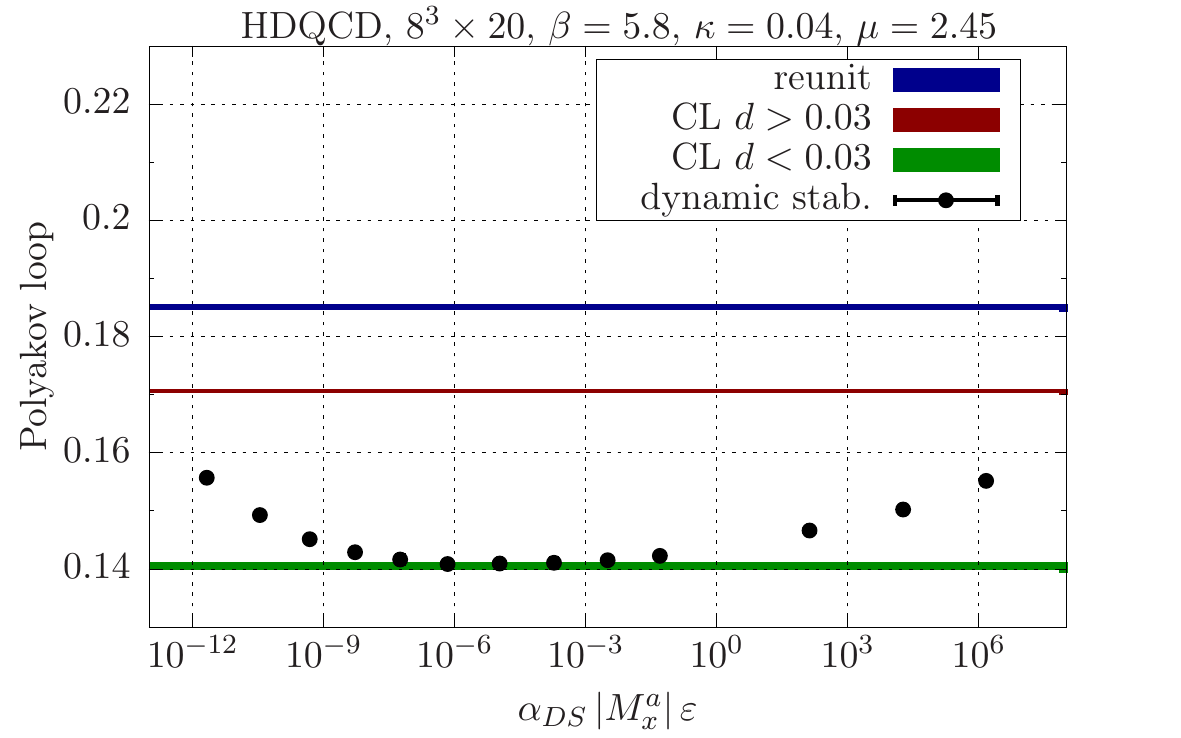}
  \caption{Expectation value of the Polyakov loop as a function of the
  control parameter $\alpha_{\mathrm{DS}}$. We also show results from simulations 
  with gauge cooling (GC) at different cutoffs for the unit. 
  norm, and with real Langevin.}
  \label{Fig:DS}
\end{figure}

The optimal choice is in the intermediate regime,
where the overall distribution of the drifts is narrower and the observables
are least sensitive to the control parameter $\alpha_{\mathrm{DS}}$. This behaviour can be
seen in Figure~\ref{Fig:DS} for the average Polyakov loop in the HDQCD model,
which will be explained below.
The figure shows results where the gauge links have been periodically reunitarised, complex Langevin simulations with gauge cooling (GC), and simulations using GC and DS.
Two separate analyses have been carried out with the data from the GC runs: in one of them all data points after thermalisation have been considered, while in the other the points after the unitarity norm reached our threshold of $d \sim 0.03$ have been disconsidered.
The GC data from the first analysis is known to be incorrect and is shown for
comparison.

Gauge cooling is known to produce reliable results when the unitarity norm is small.
The agreement with DS simulations for certain intervals of $\alpha_{\mathrm{DS}}$ 
is evidence that DS succeeds in keeping large explorations of SL($3,\mathbb{C}$) under control.
 
\section{HDQCD}\label{HDQCD}
    
The heavy dense approximation of QCD (HDQCD) is an approximation for
very large quark masses. Using the hopping parameter expansion 
and dropping all terms beyond the static limit lead to a significant
simplification of the fermion determinant: It can be written solely in terms of the Polyakov loop
$P_{\vec{x}}$ and its inverse
\begin{equation}
\det D(\mu) = \prod_{\vec{x}} \det \big( 1 + h\,\mathrm{e}^{\mu/T} \,
P_{\vec{x}} \big)^{2}\det  \big( 1 + h\,\mathrm{e}^{-\mu/T}\,
P^{-1}_{\vec{x}} \big)^2,  \quad  \text{where} \quad h = \left( 2 \, \kappa
\right)^{N_\tau}\,,
\end{equation}  
and
\begin{equation} 
	P_{\vec{x}} = \prod_{\tau=0}^{N_\tau-1} U_{(\vec{x},\tau),\hat{4}}\,,
\end{equation}
with $N_\tau$ being the lattice extent in the temporal direction.
This simplification allows for quick studies of the effect of Dynamic Stabilisation
(DS). Here, we look in particular at the distribution of the drifts, which can
be considered as a criterion for convergence~\cite{Aarts:2009uq,Aarts:2011ax}. 
\begin{figure}[htb]
  \centering
  \includegraphics[width=9cm,clip]{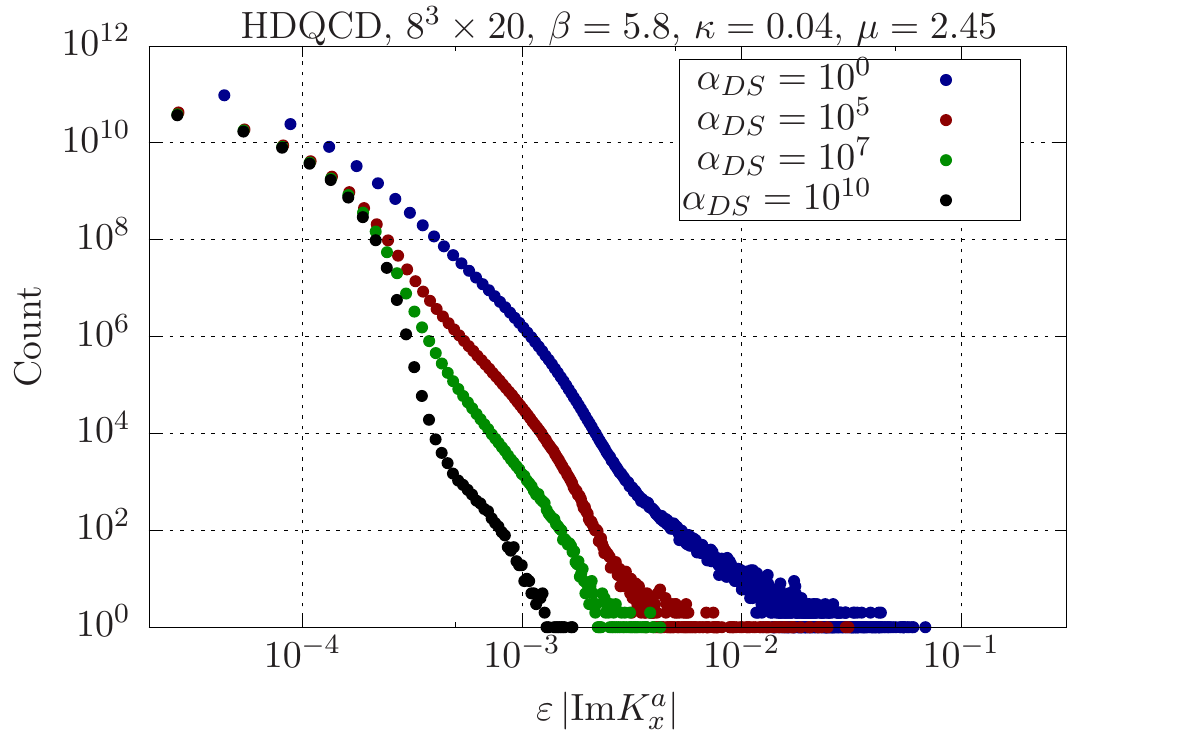}
  \caption{Histograms of the imaginary part of the drift (non-SU$(3)$
  part) as function of the magnitude $\varepsilon\, |\mathrm{Im} K^a_x|$ for four
  values of the control parameter $\alpha_{\mathrm{DS}}$. }
  \label{Fig:NonSU}
\end{figure}
Figure~\ref{Fig:NonSU} shows 
a histogram of the imaginary part of the Langevin
drift, which has been averaged over the four directions.
Larger values for $\alpha_{\mathrm{DS}}$ result in more localised
distributions for the Langevin forces and thus improving the overall convergence.
\begin{figure}[htb]
  \centering
  \includegraphics[width=9cm,clip]{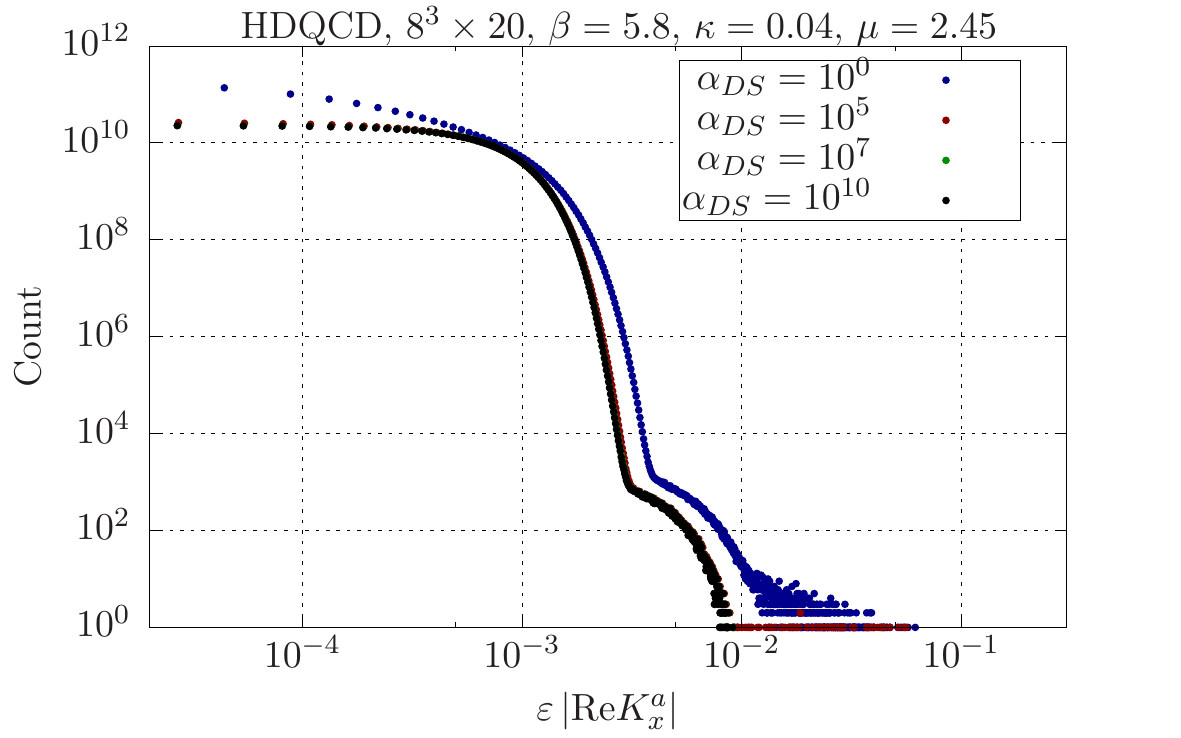}
  \caption{Histograms of the real part of the drift (SU$(3)$
  part) as function of the magnitude $\varepsilon\, |\mathrm{Re} K^a_x|$ for four
  values of the control parameter $\alpha_{\mathrm{DS}}$. } 
    \label{Fig:SU}
\end{figure}
Figure~\ref{Fig:SU} shows the equivalent distribution for the SU$(3)$
part of the Langevin drifts.  The histogram with $\alpha_{\mathrm{DS}} = 1$ is
noticeably different, which indicates that indirect effects from the non-SU$(3)$ drifts cause changes to the
distribution. For larger values, the real part of the drifts are identical and unaffected by
the additional term. If we expand the DS drift in terms of the lattice spacing,
using a na\"ive description of the gauge links 
\begin{equation}  
U_{x,\mu} = \exp \big[ \mathrm{i} \,a\, \lambda^{a} \big( A^{a}_{x,\mu} +
 \mathrm{i} \,B^{a}_{x,\mu}\big) \big],
\end{equation} 
we find that the additional drift formally vanishes in the continuum limit. This
is however not a formal proof, which is significantly harder to do, since we cannot
rewrite the DS term originating from an action principle.
\begin{figure}[thb] 
  \centering
  \includegraphics[width=9cm,clip]{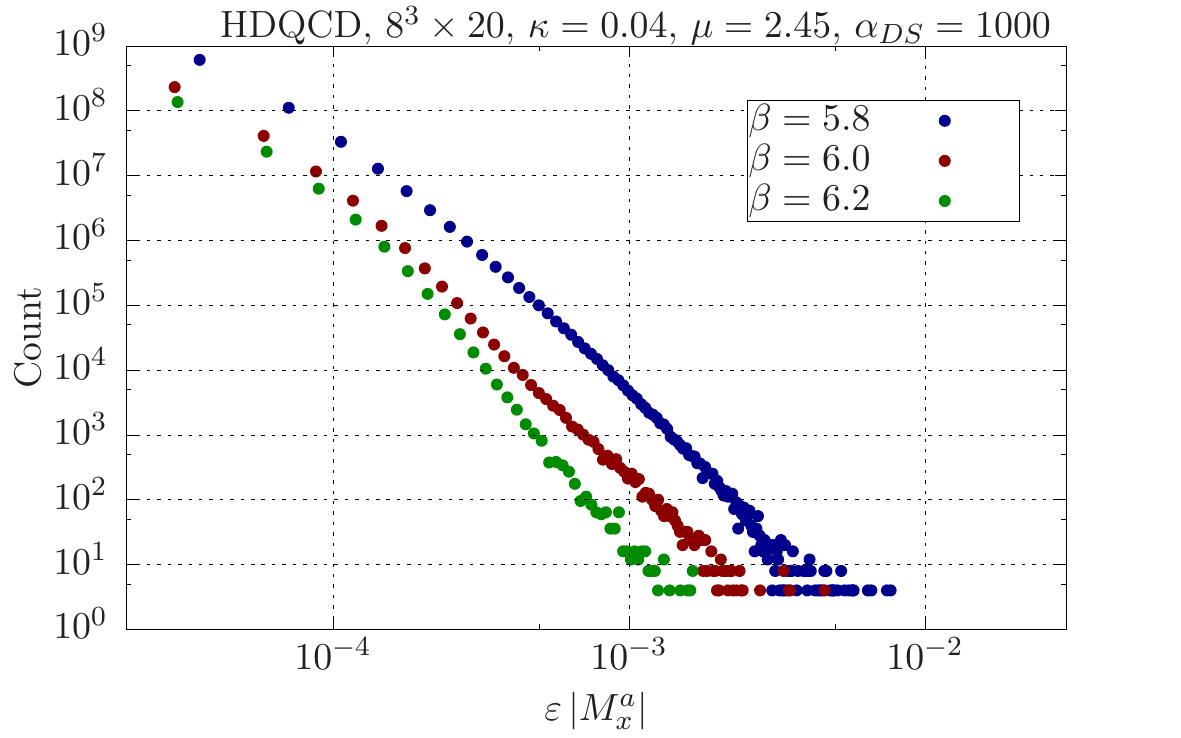}
  \caption{Histogram of the added DS force for three different values of the
  gauge coupling, keeping other simulation parameters fixed.}
  \label{Fig:Cont}
\end{figure}
Figure~\ref{Fig:Cont} shows the histograms of the DS
drift, $M_x^a$, for three values of the gauge coupling.
The counts of larger drifts can be seen to diminish as 
the continuum limit is approached, thus indicating that
 the DS force becomes trivial for large $\beta$, as desired.

\section{XY Model}\label{XY}

The three-dimensional XY model at finite density, whose action is given by
\begin{equation}
   S = - \beta \sum_{x} \sum_{\nu = 0}^{2} 
   \text{cos}\left( \phi_x - \phi_{x+\hat{\nu}} - \mathrm{i} \mu \delta_{\nu,0}
   \right)\,,
\end{equation} 
has been studied using complex Langevin simulations~\cite{Aarts:2010aq}. 
It was concluded that in the disordered phase complex Langevin simulations
fail to reproduce the correct results, which can be obtained using a dual,
sign-problem-free, world-line formulation~\cite{Chandrasekharan:2008gp,Banerjee:2010kc}. In the
following, we report on our tests of a variant of Dynamic Stabilisation applied
to the XY model\footnote{We thank Gert Aarts for suggesting this.}.
 The equivalent of the unitarity norm in this context is
simply given by the imaginary part of the complexified scalar field $\phi$,
leading to the DS force
\begin{equation}
   K_{\mathrm{DS}}  \to - D^a_{x,\mu} \, S + \textcolor{darkred}{ \mathrm{i}\,
   \alpha_{DS} \, \left[\mathrm{Im}\, \phi_x\right]^7}\,.
   \label{Eq:DSXY}
\end{equation} 
This is, however, a na\"ive implementation of a DS force, which might be not ideal.
Further studies exploring alternative definitions are ongoing. 
\begin{figure}[thb]
  \centering
  \includegraphics[width=9cm,clip]{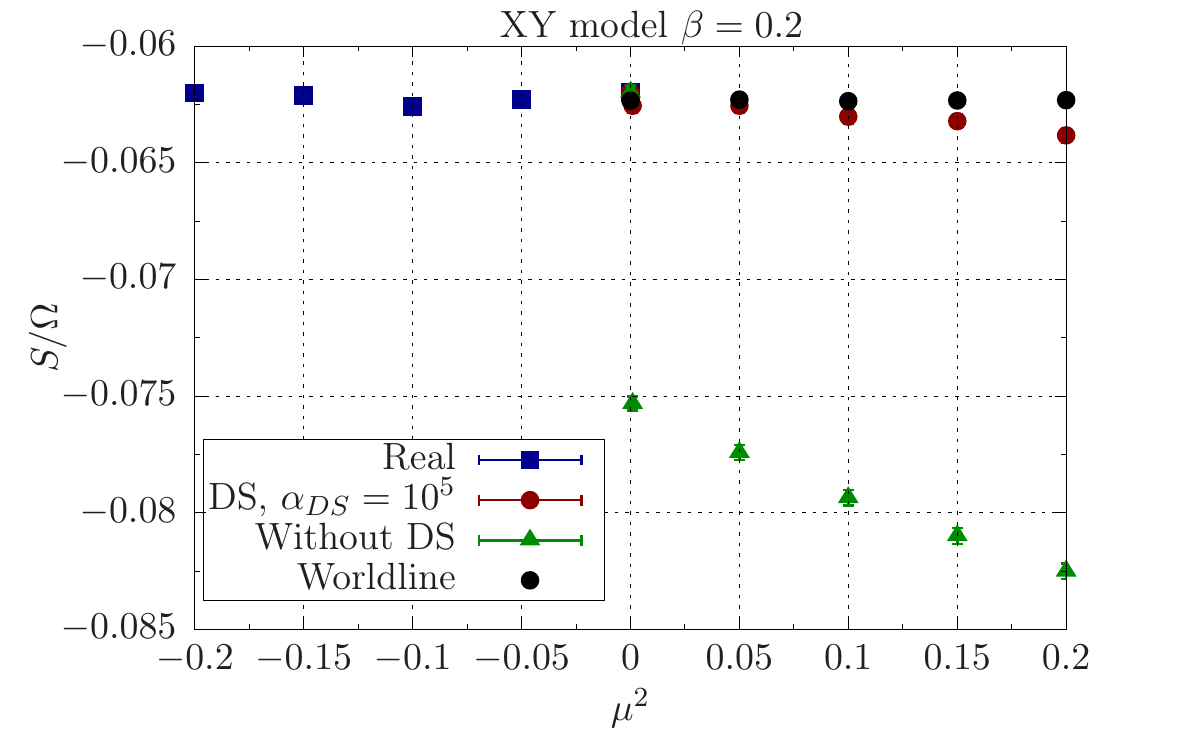}
  \caption{The action density $S/\Omega$ as a function of the chemical
  potential for the XY model in the disordered phase with $\beta = 0.2$. The
  results for the worldline formulation have been taken
  from~\cite{Aarts:2010aq}.}
  \label{Fig:XY1} 
\end{figure} Figure~\ref{Fig:XY1} shows the action density $S / \Omega$ as a function of the
chemical potential. With imaginary chemical potentials, i.e. $\mu^2 < 0$, the
theory is sign-problem-free and thus real Langevin simulations are applicable
and used. For real chemical potentials we compare simulations with and without
Dynamic Stabilisation. 
Standard complex Langevin for $\mu \neq 0$ shows a significant
deviation  from the correct results. Adding an additional drift as in
Eq.~\ref{Eq:DSXY} leads to a significant improvement, which however still shows
a small, but significant deviation from the worldline results, in particular
for larger chemical potentials.
\begin{figure}[thb]
  \centering
  \includegraphics[width=9cm,clip]{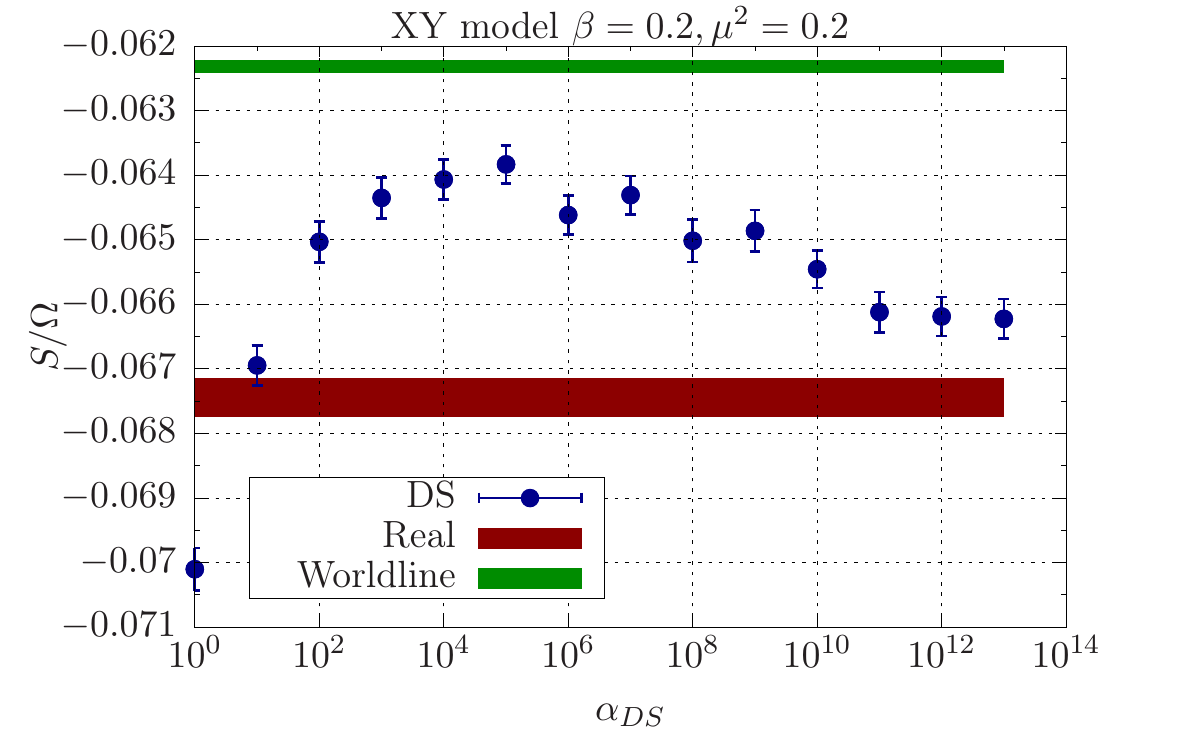}
  \caption{The action density $S/\Omega$ as a function of the control parameter
  $\alpha_{DS}$ for the XY model. The result for the worldline formulation
  (green) has been taken from~\cite{Aarts:2010aq}.}
  \label{Fig:XY2}
\end{figure}
Figure~\ref{Fig:XY2} shows a variation of the control parameter $\alpha_{DS}$
for the largest $\mu$ studied here. It is obvious that even the best choice
does not reproduce the correct result from the dual representation, shown in
green.
For very large $\alpha_{DS}$ the result approaches, as expected, the results from
real Langevin simulations, shown in red. The source of the discrepancy between
are DS and the world-line formulation, can be manifold. Our simulations are not
extrapolated in terms of the step size correction, but some first evidence
indicate that even smaller step sizes do not reconcile the observed deviation.
Further definitions in terms of the DS forces are currently being studied and other
formulations might show a better convergence.

\section{Conclusion and Outlook}\label{end}
We presented an update on applications of the method of Dynamic Stabilisation.
We studied the effects on the drifts appearing in complex Langevin
simulation of HDQCD. We find that the additional force is improving the distribution of the
drift. In particular, the non-SU$(3)$ forces becomes more localised. However, a formal justification 
for DS is still in progress. For now, the method remains a heuristic
procedure to cure instabilities occurring in complex Langevin simulations, which
have been seen in simulations with smaller gauge couplings. We further
illustrate an application of DS on the three-dimensional XY model
with finite chemical potential. We find a clear improvement towards the
correct value, which is obtainable with a sign-problem-free dual representation. 
However, DS does not solve the discrepancy entirely. Especially for larger
chemical potentials a small but significant difference remains. In future works
we will explore the possibility of improved definitions for the DS force in the XY model.

\section{Acknowledgements}\label{Acknowledgements}
We are indebted to Gert Aarts, D\'{e}nes Sexty, Erhard Seiler and Ion-Olimpiu Stamatescu
for invaluable discussions and collaboration. We are grateful for the computing
resources made available by HPC Wales. We acknowledge the STFC 
grant ST/L000369/1. B.J. was supported by the Schweizerischer Nationalfonds 
(SNF) under grant 200020-162515.  F.A. is grateful for the support through the
Brazilian government programme “Science without Borders” under scholarship number BEX 9463/13-5.


\bibliography{lattice2017}

\end{document}